\documentclass[twocolumn]{article}
\usepackage{authblk}
\usepackage{hyperref}
\usepackage{graphicx}
\usepackage{commath}
\usepackage{amssymb}
\usepackage{esint}
\usepackage[dvipsnames]{xcolor}	
\usepackage{diagbox}
\begin{document}
\title{Electroweak Monopole-antimonopole Pair in the Standard Model}
\author{Dan Zhu, Khai-Ming Wong, and Guo-Quan Wong}
\affil{School of Physics, Universiti Sains Malaysia, 11800 USM, Penang, Malaysia}
\maketitle

\begin{abstract}
We present the first numerical solution that corresponds to a pair of Cho-Maison monopole and antimonopole (MAP) in the SU(2)$\times$U(1) Weinberg-Salam (WS) theory. The monopoles are finitely separated, while each pole carries magnetic charge $\pm 4\pi/e$. The positive pole is situated in the upper hemisphere, whereas the negative pole is in the lower hemisphere. The Cho-Maison MAP was investigated for a range of Weinberg angle, $0.4675\leq\tan\theta_{\scalebox{.5}{\mbox{W}}}\leq10$, and Higgs self-coupling, $0\leq\beta\leq1.7704$. Magnetic dipole moment ($\mu_m$) and pole separation ($d_z$) of the numerical solutions are calculated and analyzed. Total energy of the system, however, is infinite due to point singularities at the locations of monopoles.
\end{abstract}

\section{Introduction}
Magnetic monopoles have remained a topic of extensive studies ever since P.A.M. Dirac \cite{Dirac} introduced the idea into Maxwell's theory. It was generalized to non-Abelian gauge theories in 1968 by Wu and Yang \cite{WuYang}. However, both Dirac and Wu-Yang monopoles possess infinite energy due to the presence of a point singularity at the origin. The first finite energy magnetic monopole solution is the 't Hooft-Polyakov monopole \cite{tHooftPolyakov} in the SU(2) Yang-Mills-Higgs (YMH) theory, found independently by G. 't Hooft and A.M. Polyakov in 1974. The mass of their monopole was estimated to be around 137 $m_{\scalebox{.5}{\mbox{W}}}$, where $m_{\scalebox{.5}{\mbox{W}}}$ is the mass of intermediate vector boson.

Since then, numerous solutions were found in the SU(2) YMH theory. In the BPS limit of vanishing Higgs potential, there exist exact monopole \cite{PrasadSommerfield} and multimonopole \cite{ExactMulti} solutions. Other well-known cases are the axially symmetric MAP of Kleihaus and Kunz \cite{KKMAP}, and the monopole-antimonopole chain (MAC) of Kleihaus et al. \cite{KKMAC}. These solutions possess finite energy and represent a chain of magnetic monopoles and antimonopoles lying in alternating order along the symmetrical axis.

In 1997, Cho and Maison \cite{ChoMaison} found a monopole solution in the standard Weinberg-Salam (WS) theory. The Cho-Maison monopole is an electroweak generalization of the Dirac monopole. It acquires a W-boson dressing and becomes a hybrid between the Dirac monopole and 't Hooft-Polyakov monopole. Moreover, its magnetic charge is twice as large because the period of the electromagnetic U(1) is $4\pi$ in the SU(2)$\times$U(1) WS theory.

The importance of Cho-Maison monopole comes from the following fact. As the electroweak generalization of the Dirac monopole, it must exist if the Standard Model is correct \cite{ChoMaison,YiSongYang,ChoMaisonMass}. One might argue that the standard WS theory cannot accommodate magnetic monopoles because the second homotopy of quotient space, SU(2)$\times$U(1)$_Y$/U(1)$_{\scalebox{0.5}{\mbox{em}}}$, is trivial. However, it is not the only monopole topology. As pointed out by Cho and Maison \cite{ChoMaison}, the WS theory, with hypercharge U(1), can be viewed as a gauged CP$^1$ model in which the normalized Higgs doublet plays the role of the CP$^1$ field. This way, the SU(2) part of WS theory has exactly the same monopole topology as the Georgi-Glashow model \cite{ChoPhilTrans}. That is, $\pi_2\left(S^2\right)=\mathbb{Z}$. Originally, the Cho-Maison monopole solution was obtained by numerical integration, but a mathematically rigorous proof of existence was established later on \cite{YiSongYang}.

The mass of Cho-Maison monopole cannot be calculated due to the point singularity at the origin. However, it is premature to deny its existence simply because it has infinite energy. Classically, the electron has an infinite electric energy but a finite mass \cite{ChoMaisonMass}. Additionally, it has been shown that the solution can be regularized. Ref. \cite{ChoMaisonMass} estimated the mass of a Cho-Maison monopole to be around 4 to 10 TeV. More recently, different methods were used, it is reported that the new BPS bound for Cho-Maison monopole may not be smaller than 2.98 TeV, more probably 3.75 TeV \cite{ChoMaisonRegularization}. Another estimate puts the lower bound of the mass of Cho-Maison monopole at 2.37 TeV \cite{ChoAdd1}.

These predictions strongly indicate that it could be produced at the LHC in the near future. So, when discovered, it will become the first magnetically-charged, topological, elementary particle. Secondly, the Cho-Maison monopole could induce the density perturbation in the early universe due to its heavy mass. For the same reason, it could become the seed of the large scale structures of the universe and the source of the intergalactic magnetic field. Moreover, it could also generate primordial magnetic black holes, which offers a possible explanation to the origin of dark matter \cite{ChoPhilTrans}. For these reasons, MoEDAL and ATLAS at the LHC are actively searching for the monopole \cite{ChoAdd3,ChoAdd4}.

Plainly, if magnetic monopoles were detected in the lab, it would be through pair production, and therefore, the importance of studying a pair of monopole and antimonopole (MAP) is self-explanatory. In 1977, Y. Nambu \cite{Nambu} predicted the existence of a pair of magnetic monopole and antimonopole bounded by a Z$^0$ flux string in the WS theory. Monopoles in an Nambu MAP carry magnetic charge $\pm4\pi\sin^2\theta_{\scalebox{.5}{\mbox{W}}}/e$. The existence of Nambu MAPs was confirmed numerically by Teh et al. \cite{Teh} using an axially symmetic magnetic ansatz. They also confirmed that Nambu MAPs are actually electroweak sphalerons reported by Kleihaus et al. \cite{KKL}.

In this work, we demonstrate that it is feasible to construct a finitely separated Cho-Maison MAP. This configuration, achieved through an axially symmetric magnetic ansatz, does not have a Z$^0$ flux string connecting the poles. Magnetic charge carried by each pole of the MAP solutions found in this study is $\pm4\pi/e$, confirming that they are indeed Cho-Maison monopoles. Additionally, it is worth noting that Gervalle and Volkov \cite{GervalleVolkov} explored Cho-Maison multimonopole solutions, which are also axially symmetric, but it is important to note that multimonopoles and MAPs are fundamentally different, since multimonopoles are just superposition of like-charges in one location.

The Cho-Maison MAP solutions were investigated at physical Weinberg angle, $\tan\theta_{\scalebox{.5}{\mbox{W}}}=0.53557042$, while the Higgs self-coupling constant, $\beta$, runs from 0 to 1.7704 and at physical $\beta=0.77818833$, while $\tan\theta_{\scalebox{.5}{\mbox{W}}}$ is allowed to vary ($0.4675\leq\tan\theta_{\scalebox{.5}{\mbox{W}}}\leq10$). The investigated quantities include magnetic dipole moment ($\mu_m$) and pole separation ($d_z$). Total energy of the configuration is infinite due to point singularities at the location of monopoles.

\section{MAP Ansatz in Standard Model}
The Lagrangian of the bosonic sector of SU(2)$\times$U(1) WS theory is given by 
	\begin{align}
	\mathcal{L}=&-\frac{1}{4}F^a_{\mu\nu}F^{a\mu\nu}-\frac{1}{4}G_{\mu\nu}G^{\mu\nu}\nonumber\\
	&-\left(\mathcal{D}_\mu\pmb\phi\right)^\dagger\left(\mathcal{D}^\mu\pmb\phi\right)-\frac{\lambda}{2}\left(\pmb\phi^\dagger\pmb\phi-\frac{\mu_{\scalebox{.5}{\mbox{H}}}^2}{\lambda}\right)^2.
	\label{eqn:Lagrangian}
	\end{align}
Here, $\mathcal{D}_\mu$ is the covariant derivative of SU(2)$\times$U(1) group and is defined as
	\begin{equation}
	\mathcal{D}_\mu=D_\mu-\frac{i}{2}g'B_\mu=\partial_\mu-\frac{i}{2}gA^a_\mu\sigma_a-\frac{i}{2}g'B_\mu,
	\end{equation}
where $D_\mu$ is the covariant derivative of SU(2) group only.\par
The SU(2) gauge coupling constant, potential and electromagnetic tensor are $g$, $A^a_\mu$ and $F^a_{\mu\nu}$. Their counterparts in U(1) gauge field are denoted as $g'$, $B_\mu$ and $G_{\mu\nu}$. The term, $\sigma_a$, is the Pauli matrices, $\pmb\phi$ and $\lambda$ are the complex scalar Higgs doublet and Higgs field self-coupling constant. Higgs boson mass and $\mu_{\scalebox{.5}{\mbox{H}}}$ are related through $m_{\scalebox{.5}{\mbox{H}}}=\sqrt2\mu_{\scalebox{.5}{\mbox{H}}}$. In addition, the Higgs field can be expressed as $\pmb\phi=H\pmb\xi/\sqrt{2}$, where $\pmb\xi$ is a column 2-vector that satisfies $\pmb\xi^\dagger\pmb\xi=1$. Metric used in this paper is ($-$+++).\par
Through Lagrangian (\ref{eqn:Lagrangian}), three equations of motion can be obtained as the following,
	\begin{align}
	\mathcal{D}^\mu\mathcal{D}_\mu\pmb\phi&=\lambda\left(\pmb\phi^\dagger\pmb\phi-\frac{\mu_{\scalebox{.5}{\mbox{H}}}^2}{\lambda}\right)\pmb\phi,\label{eqn:EoMs}\\
	D^\mu F^a_{\mu\nu}&=\frac{ig}{2}\left[\pmb\phi^\dagger\sigma^a\left(\mathcal{D}_\nu\pmb\phi\right)-\left(\mathcal{D}_\nu\pmb\phi\right)^\dagger\sigma^a\pmb\phi\right],\\
	\partial^\mu G_{\mu\nu}&=\frac{ig'}{2}\left[\pmb\phi^\dagger\left(\mathcal{D}_\nu\pmb\phi\right)-\left(\mathcal{D}_\nu\pmb\phi\right)^\dagger\pmb\phi\right].\label{eqn:EoMe}
	\end{align}

The magnetic ansatz used to obtain the Cho-Maison MAP is:
	\begin{align}
	gA^a_i=&-\frac{1}{r}\psi_1(r,\theta)\hat{n}^a_\phi\hat{\theta}_i+\frac{n}{r}\psi_2(r,\theta)\hat{n}^a_\theta\hat{\phi}_i\nonumber\\
	&+\frac{1}{r}R_1(r,\theta)\hat{n}^a_\phi\hat{r}_i-\frac{n}{r}R_2(r,\theta)\hat{n}^a_r\hat{\phi}_i,\nonumber\\
	g'B_i=&\ \frac{n}{r\sin\theta}B_s(r,\theta)\hat{\phi}_i,\;gA^a_0=g'B_0=0,\nonumber\\
	\Phi^a=&\ \Phi_1(r,\theta)\hat{n}^a_r+\Phi_2(r,\theta)\hat{n}^a_\theta=H(r,\theta)\hat{\Phi}^a.\label{eqn:Ansatz}
	\end{align}
In addition, the Higgs unit vector, $\hat{\Phi}^a$, can be written as,
	\begin{align}
	\hat{\Phi}^a=&-\pmb\xi^\dagger\sigma^a\pmb\xi\nonumber\\
	=&\cos(\alpha-\theta)\hat{n}^a_r+\sin(\alpha-\theta)\hat{n}^a_\theta=h_1\hat{n}^a_r+h_2\hat{n}^a_\theta,\nonumber\\
	\pmb\xi=&\ i
		\begin{pmatrix}
		\sin\frac{\alpha(r,\theta)}{2}e^{-in\phi}\\
		-\cos\frac{\alpha(r,\theta)}{2}
		\end{pmatrix}.
	\end{align}
The functions, $\cos\alpha$ and $\sin\alpha$ are defined as
	\begin{align}
	\cos\alpha=&\frac{\Phi_1\cos\theta-\Phi_2\sin\theta}{\sqrt{\Phi_1^2+\Phi_2^2}}=h_1\cos\theta-h_2\sin\theta,\nonumber\\
	\sin\alpha=&\frac{\Phi_1\sin\theta+\Phi_2\cos\theta}{\sqrt{\Phi_1^2+\Phi_2^2}}=h_1\sin\theta+h_2\cos\theta.
	\end{align}
Moreover, the angle, $\alpha(r,\theta)\rightarrow p\theta$ asymptotically \cite{Teh}, where $p$ is the parameter controlling the number of poles in the solution and is set to two. Finally, $H(r,\theta)=\abs{\Phi}=\sqrt{\Phi_1^2+\Phi_2^2}$ is the Higgs modulus.

In the magnetic ansatz, the spatial spherical coordinate unit vectors are defined as
	\begin{align}
	\hat{r}_i&=\sin\theta\cos\phi\,\delta_{i1}+\sin\theta\sin\phi\,\delta_{i2}+\cos\theta\,\delta_{i3},\nonumber\\
	\hat{\theta}_i&=\cos\theta\cos\phi\,\delta_{i1}+\cos\theta\sin\phi\,\delta_{i2}-\sin\theta\,\delta_{i3},\nonumber\\
	\hat{\phi}_i&=-\sin\phi\,\delta_{i1}+\cos\phi\,\delta_{i2}.
	\end{align}
Similarly, the unit vectors for isospin coordinate system are given by 
	\begin{align}
	\hat{n}^a_r&=\sin\theta\cos n\phi\,\delta^a_1+\sin\theta\sin n\phi\,\delta^a_2+\cos\theta\,\delta^a_3,\nonumber\\
	\hat{n}^a_\theta&=\cos\theta\cos n\phi\,\delta^a_1+\cos\theta\sin n\phi\,\delta^a_2-\sin\theta\,\delta^a_3,\nonumber\\
	\hat{n}^a_\phi&=-\sin n\phi\,\delta^a_1+\cos n\phi\,\delta^a_2,
	\end{align}
where $n$ is the $\phi$-winding number and is set to one in this research.

Upon substituting the magnetic ansatz (\ref{eqn:Ansatz}) into the equations of motion (\ref{eqn:EoMs}) - (\ref{eqn:EoMe}), the system of equations was reduced to seven coupled second-order partial differential equations, which were further simplified with the following substitutions,
	\begin{equation}
	x=m_{\scalebox{.5}{\mbox{W}}}r,\;\widetilde{H}=\frac{H}{H_0},\;\tan\theta_{\scalebox{.5}{\mbox{W}}}=\frac{g'}{g},\;\beta^2=\frac{\lambda}{g^2}.
	\end{equation}
Here $H_0=\sqrt2\mu_{\scalebox{.5}{\mbox{H}}}/\sqrt\lambda$ and $m_{\scalebox{.5}{\mbox{W}}}=gH_0/2$. The new radial coordinate, $x$, is dimensionless, which is then compactified through $\widetilde{x}=x/(x+1)$. The rescaled Higgs field, $\widetilde{H}$, approaches one asymptotically. Only two free parameters were left after the transformation, $\tan\theta_{\scalebox{.5}{\mbox{W}}}$ and $\beta$. Both of these parameters can be expressed in terms of the mass of elementary particles and by adopting $m_{\scalebox{.5}{\mbox{H}}}=125.10$ GeV, $m_{\scalebox{.5}{\mbox{W}}}=80.379$ GeV and $m_{\scalebox{.5}{\mbox{Z}}}=91.1876$ GeV\cite{PDG}, these parameters are calculated to be $\beta=0.77818833$ and $\tan\theta_{\scalebox{.5}{\mbox{W}}}=0.53557042$.

The seven coupled equations are then subject to the following boundary conditions. Along the positive and negative $z$-axis, when $\theta=0$ and $\pi$,
	\begin{align}
	\partial_\theta\psi_A=R_A=\partial_\theta\Phi_1=\Phi_2=\partial_\theta B_s=0,\label{eqn:BCtheta}
	\end{align}
where $A=1$, 2. Asymptotically, when $r$ approaches infinity,
	\begin{align}
	\psi_A(\infty,\theta)&=2,R_A(\infty,\theta)=B_s(\infty,\theta)=0,\nonumber\\
	\Phi_1(\infty,\theta)&=\cos\theta,\Phi_2(\infty,\theta)=\sin\theta,\label{eqn:BCrinf}
	\end{align}
and finally, at the origin, 
	\begin{align}
	&\psi_A(0,\theta)=R_A(0,\theta)=0,B_s(0,\theta)=-2,\nonumber\\
	&\Phi_1(0,\theta)\sin\theta+\Phi_2(0,\theta)\cos\theta=0,\nonumber\\
	&\partial_r(\Phi_1(r,\theta)\cos\theta-\Phi_2(r,\theta)\sin\theta)|_{r=0}=0.\label{eqn:BCr0}
	\end{align}

Using finite difference method (central difference approximation), the set of seven coupled partial differential equations was converted into a system of non-linear equations, which was then discretized onto a non-equidistant grid of $M\times N$, where $M=70$, $N=60$. The associated error with the numerical method employed is $\mathcal{O}\left(1/M^2\right)$ in $r$ direction and $\mathcal{O}\left(\pi^2/N^2\right)$ in $\theta$ direction. The region of integration covers all space which translates to $0\leq\widetilde{x}\leq1$ and $0\leq\theta\leq\pi$. Good initial guesses are needed in order for the numerical computation to converge.

\section{Properties of Cho-Maison MAP}
In the SU(2)$\times$U(1) WS model, energy density of Cho-Maison MAP solutions is obtained from the energy-momentum tensor, $T_{\mu\nu}$:
	\begin{align}
	\varepsilon=T_{00}&=\frac{1}{4}G_{ij}G_{ij}+\frac{1}{4}F^a_{ij}F^a_{ij}+\left(\mathcal{D}_i\pmb\phi\right)^\dagger\left(\mathcal{D}_i\pmb\phi\right)\nonumber\\
	&\quad\,+\frac{\lambda}{2}\left(\pmb\phi^\dagger\pmb\phi-\frac{\mu_{\scalebox{.5}{\mbox{H}}}^2}{\lambda}\right)^2.
	\end{align}
The curve of weighted energy density ($\varepsilon_{\scalebox{.5}{\mbox{W}}}=r^2\sin\theta\cdot\varepsilon$) near the $z$-axis was plotted to show that its value blows up and the singularities appear at the location of the monopoles.

To investigate the magnetic properties, the following gauge was applied:
	\begin{align}
	G&=-i
		\begin{pmatrix}
		\cos\frac{\alpha}{2}&\sin\frac{\alpha}{2}e^{-in\phi}\\
		\sin\frac{\alpha}{2}e^{in\phi}&-\cos\frac{\alpha}{2}
		\end{pmatrix}\nonumber\\
	&=\cos\frac{-\pi}{2}+i\hat{u}^a_r\sigma^a\sin\frac{-\pi}{2},\nonumber\\
	\hat{u}^a_r&=\sin\frac{\alpha}{2}\cos{n\phi}\,\delta^a_1+\sin\frac{\alpha}{2}\sin{n\phi}\,\delta^a_2+\cos\frac{\alpha}{2}\,\delta^a_3.\label{eqn:gauge}
	\end{align}
The transformed gauge potential has the following form:
	\begin{align}
	gA'^a_i=&-\frac{2n}{r}\left[\psi_2\sin\left(\theta-\frac{\alpha}{2}\right)+R_2\cos\left(\theta-\frac{\alpha}{2}\right)\right]\hat{u}^a_r\hat\phi_i\nonumber\\
	&-gA^a_i+\frac{2n\sin\frac{\alpha}{2}}{r\sin\theta}\hat{u}_\theta^a\hat\phi_i-\partial_i\alpha\;\hat{u}_\phi^a.\label{eqn:transformedA}
	\end{align}
Note here, when $a=3$, equation (\ref{eqn:transformedA}) becomes the 't Hooft gauge potential,
	\begin{align}
	gA^{'3}_i&=\frac{n}{r}\left(\psi_2h_2-R_2h_1-\frac{1-\cos\alpha}{\sin\theta}\right)\hat{\phi}_i\nonumber\\
	&=\frac{A_s}{r\sin\theta}\hat{\phi}_i.
	\end{align}

Now, as
	\begin{align}
		\begin{pmatrix}
		A_i^{\scalebox{.5}{\mbox{em}}}\\
		Z_i
		\end{pmatrix}&=
		\begin{pmatrix}
		\cos\theta_{\scalebox{.5}{\mbox{W}}}&\sin\theta_{\scalebox{.5}{\mbox{W}}}\\
		-\sin\theta_{\scalebox{.5}{\mbox{W}}}&\cos\theta_{\scalebox{.5}{\mbox{W}}}
		\end{pmatrix}
		\begin{pmatrix}
		B_i\\
		A'^3_i
		\end{pmatrix}\nonumber\\
	&=\frac{1}{\sqrt{g'^2+g^2}}
		\begin{pmatrix}
		g&g'\\
		-g'&g
		\end{pmatrix}
		\begin{pmatrix}
		B_i\\
		A'^3_i
		\end{pmatrix}\label{eqn:AZrelation},
	\end{align}
the real electromagnetic potential could be expressed as
	\begin{align}
	eA_i^{\scalebox{.5}{\mbox{em}}}&=\cos^2\theta_{\scalebox{.5}{\mbox{W}}}\left(g'B_i\right)+\sin^2\theta_{\scalebox{.5}{\mbox{W}}}\left(gA'^3_i\right),\label{eqn:Amu}
	\end{align}
where $e$ is the unit electric charge. The `em' magnetic field could then be calculated according to the mixing shown in equation (\ref{eqn:Amu}),
	\begin{align}
	B^{\scalebox{.5}{\mbox{em}}}_i&=-\frac{1}{2}\varepsilon_{ijk}F^{\scalebox{.5}{\mbox{em}}}_{jk}\nonumber\\
	&=-\frac{1}{e}\varepsilon_{ijk}\partial_j\{\cos^2\theta_{\scalebox{.5}{\mbox{W}}}B_s+\sin^2\theta_{\scalebox{.5}{\mbox{W}}}A_s\}\partial_k\phi.
	\end{align}

\begin{figure*}[t]
	\includegraphics[width=\linewidth]{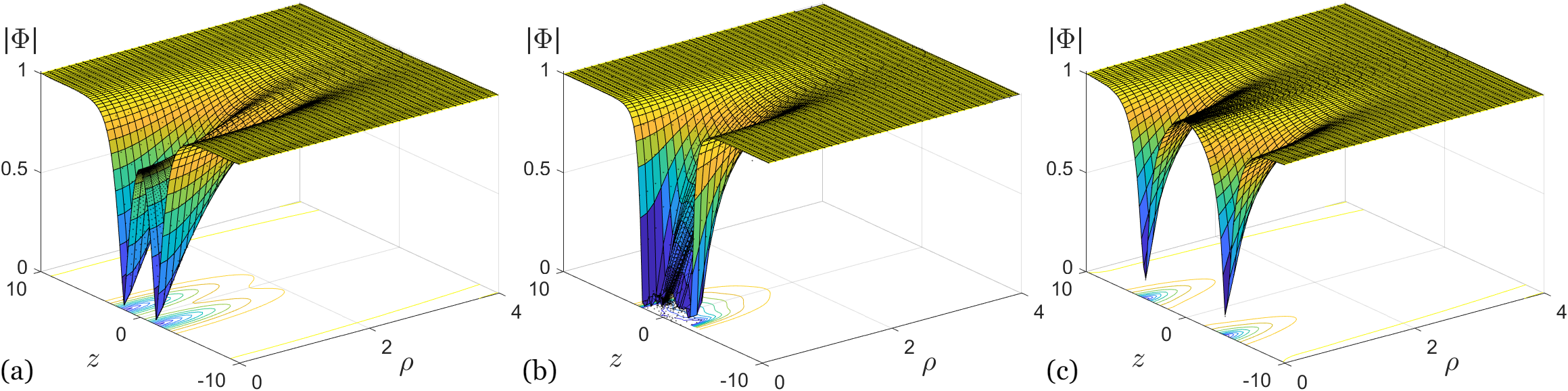}
	\caption{3D Higgs modulus plots of (a) an MAP found in the SU(2) YMH theory, (b) a Nambu MAP (electroweak sphaleron), and (c) a Cho-Maison MAP. All with $\beta=0.77818833$. For (b) and (c), $\tan\theta_{\scalebox{.5}{\mbox{W}}}=0.53557042$.}
	\label{fig:HiggsCompare}
\end{figure*}

\begin{figure}[t]
	\includegraphics[width=\linewidth]{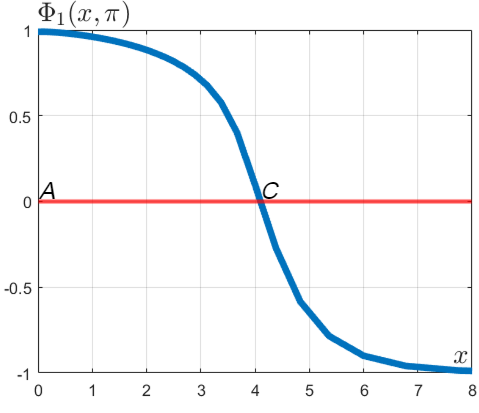}
	\caption{The curve of $\Phi_1\left(x,\pi\right)$ versus $x$ for the Cho-Maison MAP shown in Fig. \ref{fig:HiggsCompare}(c).}
	\label{fig:Phi1Curve}
\end{figure}

\begin{figure}[t]
	\includegraphics[width=\linewidth]{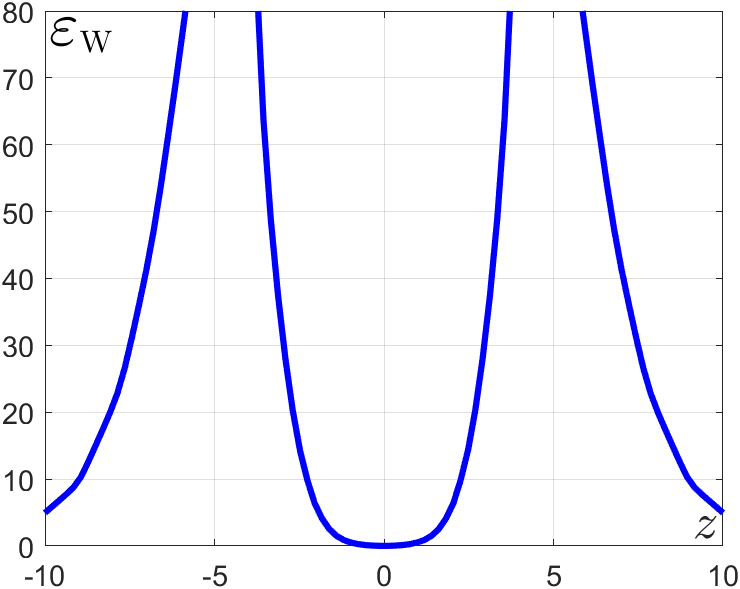}
	\caption{Plot of $\varepsilon_{\scalebox{.5}{\mbox{W}}}$ versus $z$ at $\rho=0.1077$ for a Cho-Maison MAP with $\beta=0.77818833$ and $\tan\theta_{\scalebox{.5}{\mbox{W}}}=0.53557042$. The values of $\varepsilon_{\scalebox{.5}{\mbox{W}}}$ clearly indicate that there are two point singularities at the location of monopoles.}
	\label{fig:energydensity}
\end{figure}

Through Gauss's law, the magnetic charge enclosed in a Gaussian surface, $S$, with surface element, $dS^i$, can be obtained with the following integral:
	\begin{equation}
	Q_{\scalebox{.5}{\mbox{M}}(S)}=\oiint_{\scalebox{.5}{\mbox{S}}}B_i^{\scalebox{.5}{\mbox{em}}}dS^i\label{eqn:Q_M(S)}=\iiint_{\scalebox{.5}{\mbox{V}}}\partial^iB_i^{\scalebox{.5}{\mbox{em}}}dV,
	\end{equation}
and for the magnetic charge enclosed in the upper hemisphere, the following Gaussian surface was defined:
	\begin{align}
	S_+=H^2_+\cup D^2_{\scalebox{.5}{\mbox{xy}}},
	\end{align}
where $H^2_+$ is a half sphere above the $xy$-plane and $D^2_{\scalebox{.5}{\mbox{xy}}}$ denotes a disk in the $xy$-plane centered at the origin, both $H^2_+$ and $D^2_{\scalebox{.5}{\mbox{xy}}}$ have the same radius. Taking into account the correct orientation of the surface elements, the magnetic charge enclosed can be calculated as:
	\begin{align}
	Q_{\scalebox{.5}{\mbox{M}}(S_+)}=&\iint_{H^2_+}B_i^{\scalebox{.5}{\mbox{em}}}r^2\sin\theta\,d\theta\,d\phi\;\hat r^i\nonumber\\
	&-\iint_{D^2_{\scalebox{.5}{\mbox{xy}}}}B_i^{\scalebox{.5}{\mbox{em}}}r\,dr\,d\phi\;\hat z^i.
	\end{align}
Upon applying the boundary conditions (\ref{eqn:BCtheta}) - (\ref{eqn:BCr0}), the first integral vanishes, while the second one becomes:
	\begin{align}
	\iint_{D^2_{\scalebox{.5}{\mbox{xy}}}}B_i^{\scalebox{.5}{\mbox{em}}}r\,dr\,d\phi\;\hat z^i&=\frac{2\pi}{e}\left[\cos^2\theta_{\scalebox{.5}{\mbox{W}}}\left(-2\right)+\sin^2\theta_{\scalebox{.5}{\mbox{W}}}\left(-2\right)\right]\nonumber\\
	&=-\frac{4\pi}{e}.
	\end{align}
Therefore, the magnetic charge carried by the monopole in the upper space is $Q_{\scalebox{.5}{\mbox{M}}(S_+)}=4\pi/e$, which is characteristic of a Cho-Maison monopole. Similar calculation shows that magnetic charge in the lower space is $Q_{\scalebox{.5}{\mbox{M}}(S_-)}=-4\pi/e$.

Moreover, the magnetic dipole moment, $\mu_m$, of a Cho-Maison MAP can be calculated according to the mixing shown in equation (\ref{eqn:Amu}) and considering at large $r$, $g'B_i=gA'^3_i$,
	\begin{align}
	A_i^{\scalebox{.5}{\mbox{em}}}\longrightarrow\frac{1}{e}\left(g'B_i\right)=\frac{1}{e}\frac{nB_s}{r\sin\theta}\hat{\phi}_i.
	\end{align}
Here, we perform an asymptotic expansion,
	\begin{align}
	nB_s\longrightarrow-\frac{\mu_m\sin^2\theta}{r},
	\end{align}
and therefore, $\mu_m=-nrB_s/\sin^2\theta$. The value of $\mu_m$ (in unit of $1/(e\cdot m_{\scalebox{.5}{\mbox{W}}})$) is evaluated at $\theta=\pi/2$.

\section{Results}
Figure \ref{fig:HiggsCompare} shows a comparison of 3D Higgs modulus for three MAP configurations, (a) an SU(2) MAP \cite{KKMAP}, (b) a Nambu MAP (electroweak sphaleron) \cite{Teh}, and (c) a Cho-Maison MAP. In Fig. \ref{fig:HiggsCompare}, the $\rho$-axis is defined as $\rho=\sqrt{x^2+y^2}$. Physical Higgs self-coupling, $\beta=0.77818833$, was chosen for all three solutions and physical Weinberg angle, $\tan\theta_{\scalebox{.5}{\mbox{W}}}=0.53557042$, was used for the ones shown in Fig. \ref{fig:HiggsCompare}(b) and (c). Evidently, in a Nambu MAP, the poles are connected through a Z$^0$ flux string, but in a Cho-Maison MAP, they are two separate entities, just like an SU(2) MAP.

Visually, the pole separation, $d_z$, of the Cho-Maison MAP is significantly larger than that of the MAP found in the SU(2) YMH theory. The value of $d_z$ for Cho-Maison MAP can be obtained numerically from the curve of $\Phi_1\left(x,0\right)$ and $\Phi_1\left(x,\pi\right)$. This is because $\Phi_2\left(x,\theta\right)$ is zero along the $z$-axis in boundary condition (\ref{eqn:BCtheta}) and hence, the behavior of $|\Phi|$ when $\theta=0$ or $\pi$ is solely determined by $\Phi_1$.

Figure \ref{fig:Phi1Curve} shows the curve of $\Phi_1\left(x,\pi\right)$ for the MAP solution shown in Fig. \ref{fig:HiggsCompare}(c). The magnetic antimonopole is located at $\Phi_1(x,\pi)=0$, which is labelled $C$ in Fig. \ref{fig:Phi1Curve}. On the other hand, the magnetic monopole is located at $\Phi_1(x,0)=0$ which shares the same value as that of point $C$ in Fig. \ref{fig:Phi1Curve}. The pole separation is then defined as $d_z=2\times AC$.

\begin{figure}[t]
	\includegraphics[width=\linewidth]{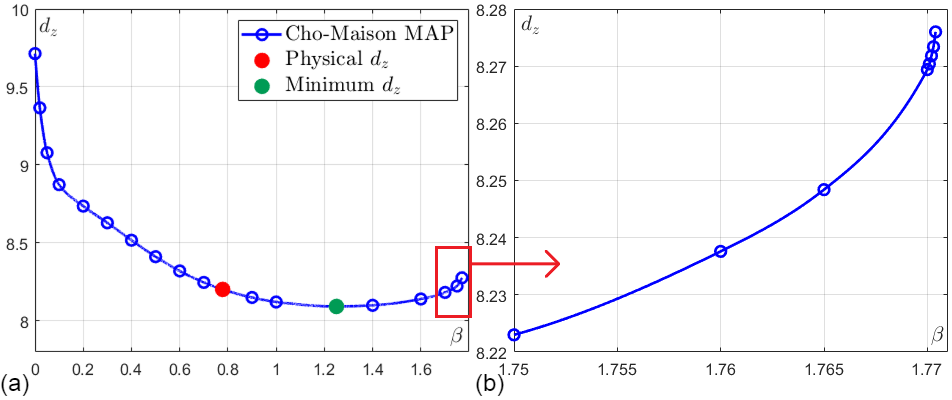}
	\caption{Plots of (a) $d_z$ versus $\beta$ for a Cho-Maison MAP with a zoomed-in version of the highlighted region shown in (b).}
	\label{fig:dz}
\end{figure}

The plot of $\varepsilon_{\scalebox{.5}{\mbox{W}}}$ versus $z$ at $\rho=0.1077$ was shown in Fig. \ref{fig:energydensity}
for the particular solution displayed in Fig. \ref{fig:HiggsCompare}(c). The weighted energy density increases rapidly near the location of monopoles, which clearly indicates there are two point singularities. As a result, total energy of the system is infinite.

By fixing Weinberg angle at $\tan\theta_{\scalebox{.5}{\mbox{W}}}=0.53557042$, the Cho-Maison MAP configuration is investigated for a range of $\beta$ from 0 to 1.7704. It is found that $d_z$ varies with $\beta$ and the plot is shown in Fig. \ref{fig:dz}(a). The value of $d_z$ starts off as 9.7140 when $\beta=0$, then monotonically decreases until $\beta_{\scalebox{.5}{\mbox{min}}}=1.25$, where the local minimum $d_z=8.0920$ (green dot) is reached. For physical $\beta=0.77818833$, $d_z$ is measured to be 8.2012 (red dot). Instead of reaching a constant value, $d_z$ increases after $\beta_{\scalebox{.5}{\mbox{min}}}=1.25$ until $\beta_{\scalebox{.5}{\mbox{c}}}=1.7704$, where no solution can be found for $\beta>\beta_{\scalebox{.5}{\mbox{c}}}$. The existence of an upperbound in $\beta$ is unexpected as SU(2) MAPs do not possess such a feature. The corresponding value for $d_z$ when $\beta=\beta_{\scalebox{.5}{\mbox{c}}}$ is 8.2760 and the behavior of $d_z$ near $\beta_{\scalebox{.5}{\mbox{c}}}$ is shown in Fig. \ref{fig:dz}(b). It can be seen that gradient of the curve drastically increases after $\beta=1.765$. 	

\begin{figure}[t]
	\includegraphics[width=\linewidth]{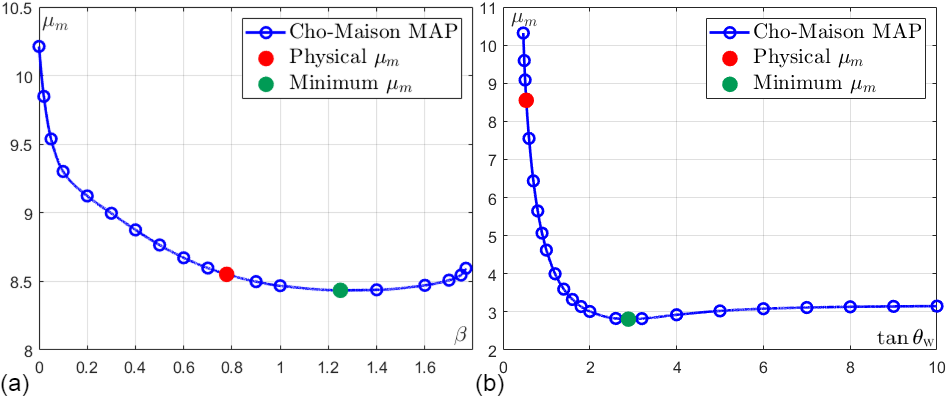}
	\caption{Plots of (a) $\mu_m$ versus $\beta$ when $\tan\theta_{\scalebox{.5}{\mbox{W}}}=0.53557042$, (b) $\mu_m$ versus $\tan\theta_{\scalebox{.5}{\mbox{W}}}$ when $\beta=0.77818833$ for a Cho-Maison MAP.}
	\label{fig:mum}
\end{figure}

\begin{table*}
	\caption{Table of $\mu_m$ and $d_z$ for selected $\beta$ of Cho-Maison MAP solutions at physical Weinberg angle $\tan\theta_{\scalebox{.5}{\mbox{W}}}=0.53557042$ (Numbers in the table are rounded).}
	\centering
	\begin{tabular*}{\textwidth}{@{\extracolsep{\fill}}c c c c c c c c c c c c}
		\hline
		$\beta$	&0		&0.2		&0.4		&0.6		&\color{red}{0.7782}&0.9		&1		&\color{ForestGreen}{1.25}		&1.4		&1.6		&1.7704\\
		\hline
		$\mu_m$	&10.2122	&9.1227	&8.8754	&8.6719	&\color{red}{8.5514}	&8.4983	&8.4684	&\color{ForestGreen}{8.4350}	&8.4385	&8.4715	&8.5958\\
		$d_z$		&9.7140	&8.7356	&8.5170	&8.3198	&\color{red}{8.2012}	&8.1496	&8.1208	&\color{ForestGreen}{8.0920}	&8.0994	&8.1398	&8.2760\\
		\hline
	\end{tabular*}
	\label{table:1}
\end{table*}

\begin{table*}
	\caption{Table of $\mu_m$ and $d_z$ for selected $\tan\theta_{\scalebox{.5}{\mbox{W}}}$ of Cho-Maison MAP solutions at physical Higgs self-coupling constant $\beta=0.77818833$ (Numbers in the table are rounded).}
	\centering
	\begin{tabular*}{\textwidth}{@{\extracolsep{\fill}}c c c c c c c c c c c c}
		\hline
		$\tan\theta_{\scalebox{.5}{\mbox{W}}}$	&0.4675	&\color{red}{0.5356}&0.6		&0.8		&1		&2		&\color{ForestGreen}{2.89}		&4		&6		&8		&10\\
		\hline
		$\mu_m$	&10.3178	&\color{red}{8.5514}	&7.5528	&5.6489	&4.6219	&3.0113	&\color{ForestGreen}{2.8098	}&2.9240	&3.0831	&3.1329	&3.1513\\
		$d_z$		&9.9384	&\color{red}{8.2012}	&7.2484	&5.4582	&4.4912	&2.8624	&\color{ForestGreen}{2.6120}&2.7568	&2.9584	&3.0228	&3.0470\\
		\hline
	\end{tabular*}
	\label{table:2}
\end{table*} 

The plot of $\mu_m$ versus $\beta$ for a Cho-Maison MAP is shown in Fig. \ref{fig:mum}(a). The shape of the curve is basically identical to Fig. \ref{fig:dz}(a), which is expected, considering the close relations between $\mu_m$ and $d_z$. The physical value is measured to be $\mu_m=8.5514$ (red dot), while the minimum $\mu_m$ is 8.4350 around $\beta_{\scalebox{.5}{\mbox{min}}}=1.25$ (green dot). We also investigate the behavior of $\mu_m$ by fixing $\beta=0.77818833$, while allowing $\tan\theta_{\scalebox{.5}{\mbox{W}}}$ to vary between 0.4675 and 10, Fig. \ref{fig:mum}(b). It is found that $\mu_m\rightarrow10.3178$ as $\tan\theta_{\scalebox{.5}{\mbox{W}}}\rightarrow0.4675$, where a lower bound is reached, below which no solutions can be found. In a similar manner, $\mu_m$ decreases until $\tan\theta_{\scalebox{.5}{\mbox{W}}}=2.89$, then increases slightly with increasing $\tan\theta_{\scalebox{.5}{\mbox{W}}}$ before converging to a limiting value. The minimum value found this way is much lower, $\mu_m=2.8098$ (green dot). Selected data of Cho-Maison MAP is tabulated in Table \ref{table:1} and \ref{table:2}.

\section{Conclusion}
In conclusion, we have found numerical solutions in the SU(2)$\times$U(1) WS theory corresponding to a pair of Cho-Maison monopole and antimonopole. Poles of this MAP configuration are not connected through a Z$^0$ flux string and each pole carries magnetic charge $\pm4\pi/e$. The Cho-Maison MAP resides in the topological trivial sector, indicating its possible existence in nature, but its life-time or the interactions between constituents of this MAP is yet to be answered.

When $\tan\theta_{\scalebox{.5}{\mbox{W}}}$ is fixed at 0.53557042, there exists an upperbound for the Cho-Maison MAP at $\beta_{\scalebox{.5}{\mbox{c}}}=1.7704$, beyond which no solutions can be found. This is a feature that MAP solutions found in the SU(2) YMH theory do not possess. When $\beta$ is fixed at 0.77818833, a lower bound exists at $\tan\theta_{\scalebox{.5}{\mbox{W}}}=0.4675$, where both $\mu_m$ and $d_z$ reach their maximum values of 10.3178 and 9.9384. However, as solutions in the SU(2)$\times$U(1) WS theory are controlled by two parameters, $\beta$ and $\tan\theta_{\scalebox{.5}{\mbox{W}}}$, finding global extrema must be done on a 2-dimensional curve, like $d_z\left(\beta,\tan\theta_{\scalebox{.5}{\mbox{W}}}\right)$. Lastly, we were able to accurately measure the physical values of $\mu_m$ and $d_z$ when $\beta=0.77818833$ and $\tan\theta_{\scalebox{.5}{\mbox{W}}}=0.53557042$. The magnetic dipole moment of a Cho-Maison MAP is $\mu_m=8.5514$ in unit of $1/(e\cdot m_{\scalebox{.5}{\mbox{W}}})$ and the pole separation is $d_z=8.2012$ in unit of $1/m_{\scalebox{.5}{\mbox{W}}}$.

Although the numerical method employed in this research is a direct generalization of a well-known procedure used in Refs. \cite{KKMAP,KKL}, 
the solutions obtained in this study are unique and new. Indeed, the originality of this work lies in the actual construction of a novel numerical solution, which is a finitely separated Cho-Maison MAP.

Finally, due to the presence of point singularities at the locations of monopoles, total energy of the system is infinite. This also prohibits us from determining if the solutions are bound states. However, it is possible to regularize the solutions by introducing a non-trivial U(1) hypercharge permeability in the form of a dimensionless function, $\epsilon\left(\pmb\phi\right)=\left(H/H_0\right)^8$ \cite{ChoMaisonMass,ChoMaisonRegularization}. With this, the total energy of Cho-Maison MAP can be evaluated. This will be reported in a future work. Additionally, electric charges can be introduced into the system forming a pair of dyon and antidyon.


\end{document}